\documentclass[useAMS,usenatbib,onecolumn,usenatbib]{mnras}

\usepackage[utf8]{inputenc}         
\usepackage[english]{babel}

\usepackage{graphicx}
\usepackage{longtable}
\usepackage{amssymb} 
\graphicspath{{img/}} 

\def\stella{{\sc Stella}}
\def\mdlA{\texttt{R10M3ht3t3E3}}

\title[MASTER OT J004207.99+405501.1/M31LRN 2015 Luminous Red Nova in M31]
{\large{MASTER OT J004207.99+405501.1/M31LRN 2015 Luminous Red Nova in M31: Discovery, Light Curve, Hydrodynamics,  Evolution}}
\author[V. M. Lipunov et al.]{\small{V. M. Lipunov$^{1,2}$ \thanks{E-mail: $lipunov2007@gmail.com$},S.Blinnikov$^{2,3}$, E.Gorbovskoy$^{1,2}$,  A.Tutukov$^{4}$, P.Baklanov$^{3,5,6}$, V.Krushinski$^{7}$,  N. Tiurina$^{2}$},
\newauthor \small{P. Balanutsa  $^{2}$, A. Kuznetsov  $^{2}$, V. Kornilov$^{1,2}$, I. Gorbunov$^{2}$, V.Shumkov$^{2}$, V.Vladimirov$^{2}$,  O. Gress$^{8}$, N. M. Budnev  $^{8}$, K. Ivanov $^{8}$},
 \newauthor \small{A. Tlatov$^{9}$, I.Zalozhnykh $^{4}$, Yu. Sergienko $^{10}$, A. Gabovich  $^{10}$, V. Yurkov$^{10}$}\\
$^{1}$M.V.Lomonosov Moscow State University, Faculty of Physics, Leninskie gory, GSP-1, Moscow, 119991, Russia \\
$^{2}$M.V.Lomonosov Moscow State University, Sternberg Astronomical Institute, Universitetsky~pr.,~13,~Moscow, 119234, Russia\\
$^{3}$ Alikhanov Institute for Theoretical and Experimental Physics  B.~Cheremushkinskaya,~25,~Moscow, 117218 Russia,\\
       Kavli IPMU (WPI), Kashiwa, Japan, \\
$^{4}$ Institute of Astronomy of Russian Academy of Science Pyatnitskaya str., 48, Moscow 119017 Russia \\
$^{5}$ Novosibirsk State University, Novosibirsk 630090, Russia \\
$^{6}$ National Research Nuclear University (MEPhI), 115409 Moscow, Russia \\
$^{7}$ Kourovka Astronomical Observatory, Ural Federal University, Lenin ave. 51, Ekaterinburg 620000, Russia\\
$^{8}$ Applied Physics Institute, Irkutsk State University, 20, Gagarin blvd,664003, Irkutsk, Russia \\
$^{9}$ Kislovodsk Solar Station of the Main (Pulkovo) Observatory RAS, P.O.Box 45, ul. Gagarina 100, Kislovodsk 357700, Russia\\
$^{10}$ Blagoveschensk State Pedagogical University, Lenin str., 104, Amur Region, Blagoveschensk 675000, Russia\\}

\begin{document}

\date{Accepted 2017 XXX. Received 2017 Apr; in original form 2016 Feb}
\pagerange{\pageref{firstpage}--\pageref{lastpage}} \pubyear{2016}
\maketitle
\label{firstpage}

\begin{abstract}
\footnotesize{We report the discovery and multicolor (VRIW) photometry of a rare explosive star MASTER OT J004207.99+405501.1 - a luminous red nova - in the Andromeda galaxy M31N2015-01a. We use our original light curve acquired with identical MASTER Global Robotic Net telescopes  in one photometric system: VRI during first 30 days and W (unfiltered) during 70 days. Also we added publishied multicolor photometry data  to estimate the mass and energy of the ejected shell, and discuss the likely formation scenarios of outbursts of this type.  We propose the interpretation of the explosion, that is consistent with the  evolutionary scenario where star merger is a natural stage of the evolution of close-mass stars and may serve as an extra channel for the formation of nova outbursts.}
\end{abstract}


\section{Introduction}
The optical transient MASTER OT J004207.99+405501.1 ~\citep{Shumkov2015} discovered by the MASTER global robotic telescope network ~\citep{Lipunov2010} was found to belong to a rare type of luminous red novae \citep[LRNe,][]{Kurtenkov2015a,Kurtenkov2015b,Williams2015} whose history began with the discovery of the outburst of M31-RV by \citet{Rich1989}.
However, the canonical prototype of this class of events is now considered to be the
outburst of the star V838 Monocerotis ~\citep{Munari2002a}, which reached an absolute
magnitude of -10 at maximum light ~\citep{Munari2002b}.
Luminous red novae differ from
common members of this class primarily by the apparent lack of  any thermonuclear processes,
their large emitted energy, a characteristic plateau on the light curve \citep{Ivanova2013}, and very strong reddening that varies with time.
The plateau phase indicates that compared to common novae, LRNe have more massive and
dense envelopes where a stationary recombination front forms that has approximately
constant luminosity like that which occur during the explosions of type IIP supernovae \citep{MacLeod2016}.
The observation of the progenitor of V1309 Sco by \citet{Tylenda2011}, who found it to be a
contact binary with a period of 1.4 days, provided strong support for the merging mechanism
of red nova outbursts.

All these results made the object extremely popular among observers operating on all
sorts of instruments - from one-meter telescopes to the Spitzer infrared space telescope
and SWIFT gamma-ray laboratory ~\citep{Bersier2015,Dong2015,Fabrika2015,Harmanen2015,Hodgkin2015,Adams2015a,Adams2015b,Kurtenkov2015a,Kurtenkov2015b,Ovcharov2015,Pessev2015a,Pessev2015b,Pessev2015c,Pessev2015d,Geier2015,Shumkov2015b,Srivastava2015,Steele2015,Rich1989,
Wagner2015,Williams2015}. 
\citet{Dong2015}  have identified the likely progenitor in archival SDSS, CFHT, Local
Group Survey and HST imaging data. Several papers have already been published reporting
detailed observations and interpretation of the outburst of MASTER OT
J004207.99+405501.1 / M31N 2015-01a, from which we would like to mark \citet{Williams2015} with spectroscopic and photometric observations obtained by the
Liverpool Telescope, that showed the LRN becoming extremely red, and where the authors
discuss the possible progenitor scenarios of this system.

In this paper we report 
the earliest photometric observations, and the complete light curve acquired with
identical telescopes of the MASTER global network. 
We use this light curve to estimate the mass of the ejected envelope, and discuss the
formation scenario of such outbursts based on our experience in the population synthesis of
binary stars ~\citep{Tutukov1981,Masevich1988,Lipunov1996} and propose the interpretation of
the explosion.

\section{THE DISCOVERY OF MASTER OT J004207.99+405501.1 $/$ M31N2015-01a}
MASTER OT J004207.99+405501.1 $/$ M31N2015-01a was discovered by the MASTER auto-detection system  ~\citep{Lipunov2010} during the survey performed by MASTER-Kislovodsk observatory on 2015-01-13.63235 UT  \citep{Shumkov2015}.

\begin{figure}
\centering
\center{\includegraphics[width=1\linewidth]{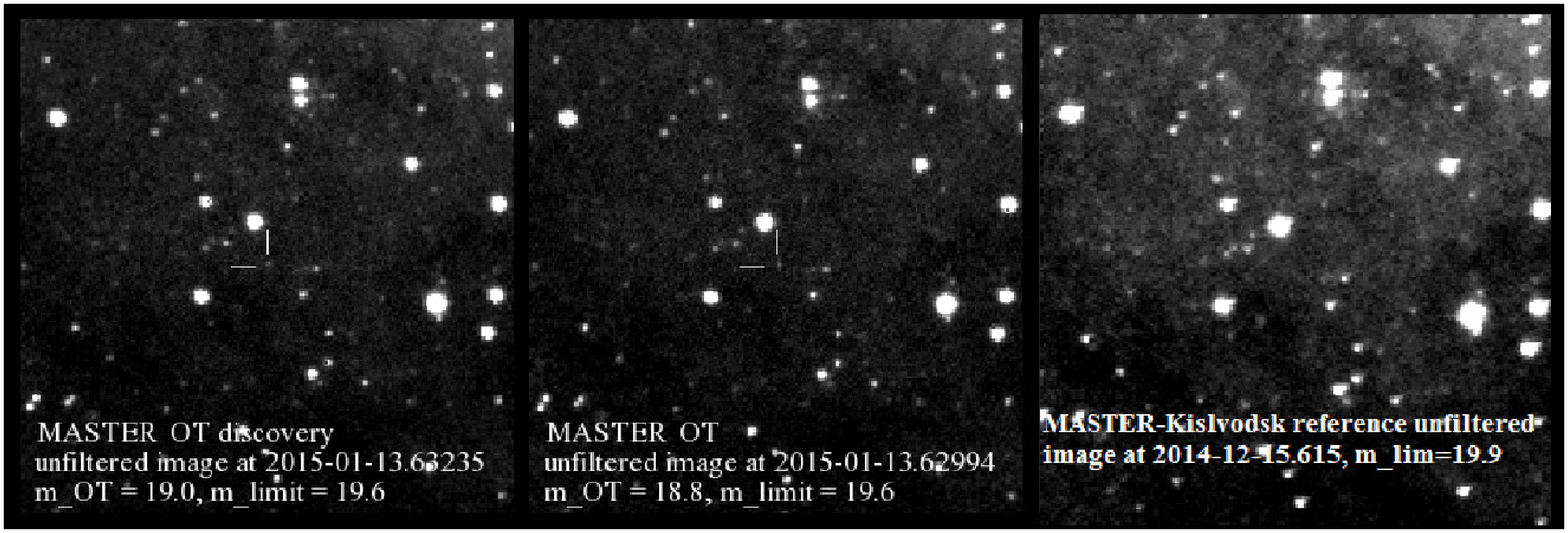}}
\caption{The discovery MASTER image of MASTER OT J004207.99+405501.1/ M31LRN . The third right one is the reference MASTER-Kislovodsk image without transient}
\label{fig:RedNova0}
\end{figure}

A total of four images containing this optical transient were acquired with the unfiltered limiting magnitudes of $m_{\rm OT}=19.0$ and $19.6$ (Fig.~\ref{fig:RedNova0}), and a
reference image with 21.1 unfiltered limiting magnitude was selected from the MASTER-Kislovodsk database.
MASTER-Kislovodsk is a node of the MASTER Global Robotic Net\footnote{\tt http://observ.pereplet.ru}.
All MASTER telescopes have  identical optical schemes and are equipped with identical sets of polarization and BVRI filters  \citep{kornilov,Pruzhinskaya2014}.

The main advantages of MASTER instruments are the following: (1) wide 8 square degree
(twin $2.05^{\circ}\times2.05^{\circ}$) field of view of the main MASTER-II optical
channel (with a limiting unfiltered magnitude of up to 20-21 per 60-180s exposition)
<and an even bigger 800 square degree (twin $16x24^{\circ}$) field of view of Very Wide
Field cameras, i.e. MASTER-VWFC (with a limiting magnitude of up to 11-12m and 13.5-15m
for 1-s and coadded images, respectively); (2) twin tubes that can be pointed to
different fields (allowing wide FERMI error-boxes to be observed almost in real-time)
and used to observe the frame in different polarizations and in BVRI filters ~\citep{Lipunov2010,kornilov,Lipunov2007,Gorbovskoy2013}.

The main goal of MASTER network is to detect prompt GRB emission by providing rapid
response to GRB-alerts . MASTER has a very fast positioning system  that makes it very
suitable for follow-up programs  such as prompt optical observations of GRB neutrinos
and GW alerts  etc. When not engaged in alert-triggered observations MASTER carries out
sky survey programs including the Andromeda survey in order to discover optical
transients of different nature (more than 10 types) and to investigate all most
important problems of the modern astrophysics.

A unique key feature of MASTER is our software that provides full information about all
optical sources detected on every image one to two minutes after the CCD readout. This
information includes the full classification of all sources found in the image, the
data from previous MASTER-Net archived images for every source, full information from
VIZIER database and from all open sources (e.g., Minor planet mpchecker center),
computed orbital elements for moving objects, etc. In search tasks a real astrophysical
source cannot occupy only 1, 2 or 4 pixels, because such objects can never be proved to
be real rather than artifacts. A real transient must occupy more than 10 pixels and
exhibit a specific profile to be distinguished from a clump of several hot-pixels .

MASTER own software discovers optical transients not just by analyzing the difference
between the previous and current images, but also by fully identifying of every source
at every image. This MASTER software allowed us to discover more then 1200 (up to
September 2016) optical transients in a fully automatic mode\footnote{\tt http://observ.pereplet.ru/MASTER\_OT.html}.
We also faced a challenging task of discovering all sources seen against the Andromeda
disc structure, especially during the rising stage. We solved this problem before
January 2015 and started our  nova search survey in the Andromeda galaxy.
At the end of 2014 the global MASTER network of twin robotic telescopes ~\citep{Lipunov2010} started automatic search for optical transients in the Andromeda galaxy.

MASTER has been observing the Andromeda galaxy every night, weather permitting, resulting in thousands of frames available for accurate photometry.

\section{Observations and data reduction}

Monitoring has been carried out with MASTER network telescopes for 72 days after our discovery of this LRN. We acquired a total of about 400 white-light frames and 130 frames with V, R, and I-band filters with 180-s exposures (see Tables 1,2,3 with VRIW-photometry). All observations passed in automatic mode. Thereby, some of the frames were acquired through light clouds.

For calibration we used the dark frames, acquired on the evening before observations, and twilight flats. Calibration, frame clipping, and astrometric reduction was performed automatically on each observatory of the network. For our photometry we used the $15\times15\, \mbox{arcmin}^2$  frame area centered on the object with about 60 comparison stars from magnitude 13 to 17 in the V band.


We used IRAF/apphot package  to perform photometry with an optimal aperture for each frame \citep{Tody1993}.
The resulting instrumental magnitudes were corrected using the Astrokit tool to minimize the standard deviation for the ensemble of comparison stars \citep{Burdanov14}.

For transformation of the instrumental magnitudes to a standard system, we used 56 nearby stars from UCAC4 catalog \citep{Zacharias2013}.
R and I magnitudes calculated from UCAC4 r- and i-bands by the following equation from \cite{Lupton2005}:

$I = r - 1.2444\cdot(r - i) - 0.3820$

$R = r - 0.2936\cdot(r - i) - 0.1439$

Some of comparison stars are presented in LGGS catalog \citep{Massey2006}, 20 and 8 for R- and I-band respectively. We used this information for checking of calculated magnitudes. Median of residual between calculated magnitudes and data from LGGS is 0.046m for R and 0.035m for I band. White light magnitudes calculated from UCAC4 B and calculated R by the equation 

$W=0.2\cdot B+0.8\cdot R$


Then, we cleared our data from the  points with big error and deviation. Visual inspection of appropriate frames showed that they were obtained through the light clouds, and in two cases the problem was caused by cosmic ray particles. Finally, we binned data points of each night by calculation of mean. Error for binned point calculated as standard deviation.

We converted our apparent magnitudes to absolute magnitudes with the adopted distance modulus of 24.43 (Feedman, Madore 1990).

Complete binned by night light curve of this Nova obtained with MASTER Global Network (MASTER-Kislovodsk, MASTER-Tunka, and MASTER-Ural) is presented at Figure 2. The data are given in the Appendix (Tables 2,3,4,5).

M31 has a distance modulus of $(m-M)_0 = 24.43 \pm 0.06$. Given the adopted foreground reddening of $ E_{(B-V)} = 0.062$ \citep{Schlegel1998} and maximum total extinction at that position in M31 of $E_{(B-V)} = 0.18$  \citep{Montalto2009}, we conservatively assume that M31LRN was subject to the reddening of $E_{(B-V)} = 0.12 \pm 0.06$, implying an absolute peak magnitude of $M_V = -9.4 \pm 0.2 $ for the transient.


\begin{figure}
\centering
	\center{\includegraphics[width=1\linewidth]{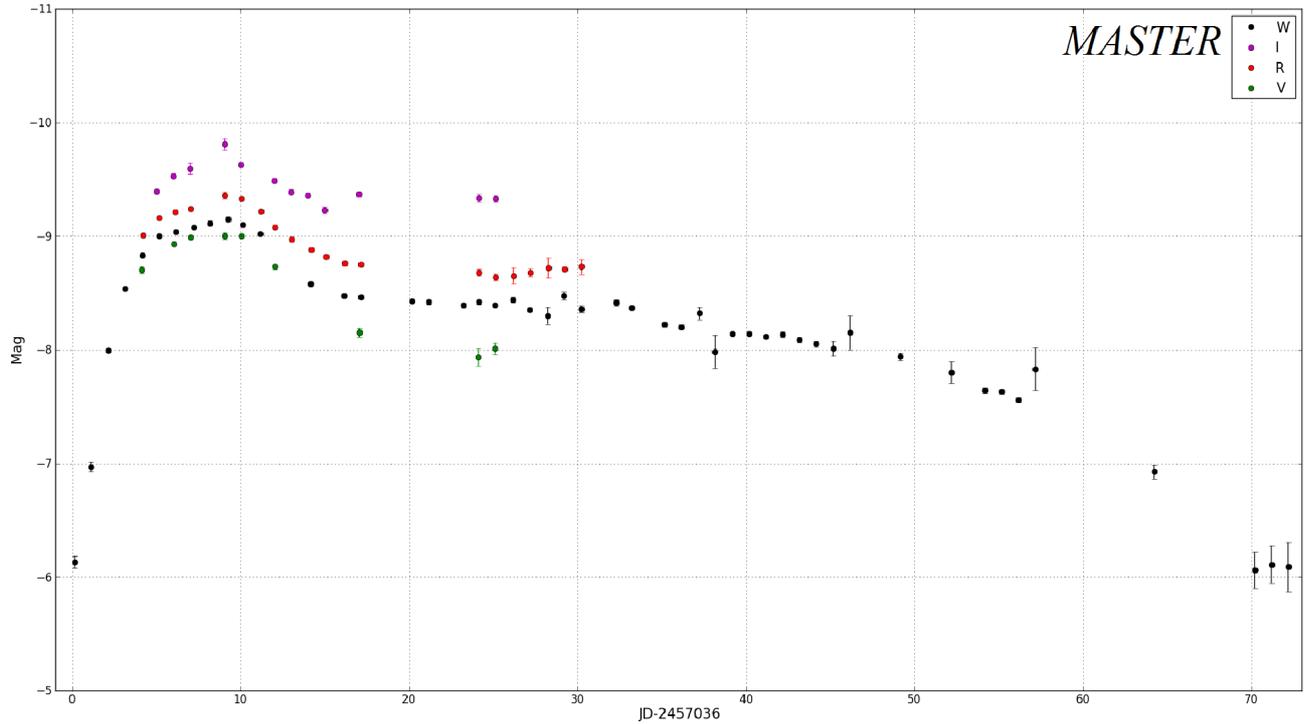}}
\caption{Complete light curve of the M31LRN Luminous Red Nova (MASTER OT J004207.99+405501.1) obtained with MASTER-Kislovodsk and MASTER-Tunka(the same instuments, i.e. the same photometry system). The adopted distance modulus is 24.43 \citep{Freedman1990}. The left-hand axis gives the apparent magnitude scale. The data are given in the Tables 2,3,4.} 
\label{fig:AveragedLC}
\end{figure}


 \section{Physics in the Model}

At present, there is no sure answer to the question: ``What is the mechanism responsible for the energy release in LRNe?''.
It may be:
\begin{enumerate}
 \item stellar mergers \citep{Soker2003,Tylenda2011}
 \item an unusual SN mechanism \citep{Lovegrove2013}
 \item a classical nova mechanism \citep{Shara2010}
 \item giant planet capture \citep{Retter2003}
 \item extreme AGB stars \citep{Thompson2009}.
\end{enumerate}

In our  study of M31LRN  hundreds of models have been tested for simulating nova
explosions in a close binary system with a common envelope with various initial parameters (in spherically symmetric approximation).
The parameters of the most suitable models used for the analysis of processes in the LRN, are shown in Table~\ref{tab:models}.
For modelling M31LRN we have adapted the code \stella\ \citep{Blinnikov2006},which is widely used for
computing supernova explosions  \citep{Woosley2007,Baklanov2015,Tolstov2015}.
\stella\ code is a set of programs for multi-group radiative hydrodynamics that can be used to compute the light
curves of supernovae of various types in the spherically symmetric approximation with the
thermodynamics of stellar plasma treated in LTE approximation. The code can also be used to
compute nova explosions, although in this case certain adjustments are needed because of the
lower velocity gradients in novae compared to supernovae and because emission in lines has in
some cases to be computed beyond the framework of Sobolev approximation. We constructed the
initial models in the same way as described in \citep{Baklanov2005}.

\begin{table}
\caption{Parameters of the model runs}
\label{tab:models}
\begin{tabular}{lccccc}
\hline
\bf{Name} &  $\mathbf{M_{total}}$ & $\mathbf{M_{heat}}$ & $\mathbf{R_{initial}}$ & $\mathbf{E_{exp}}$  & $\mathbf{t_{\rm heat}}$ \\
& $M_{\odot}$ & $M_{\odot}$ & $R_{\odot}$ &  $10^{48}$~ergs & s \\
\hline
\mdlA &  3  & 3  & 10 &  3 & $10^4$ \\ 
R10M3ht02t3E4 &  3  & 0.2  & 10 &  4 & $10^3$ \\ 
R1.6M2E008 &  2  & 2  & 1.6 &  8 & $10^4$ \\ 
R5\_M3\_E003   &  3  & 3  & 5 &  3 & $10^4$ \\ 
R20\_M3\_E003   &  3  & 3  & 20 &  3 & $10^4$ \\ 
R10\_M2\_E003   &  2  & 2  & 10 &  3 & $10^4$ \\ 
R10\_M5\_E003   &  5  & 3  & 10 &  3 & $10^4$ \\ 
R10\_M3\_E001   &  3  & 3  & 10 &  1 & $10^4$ \\ 
R10\_M3\_E008   &  3  & 3  & 10 &  8 & $10^4$ \\ 
\hline
\end{tabular}
\end{table}

\begin{figure}
  \centering
  \includegraphics[width=\linewidth]{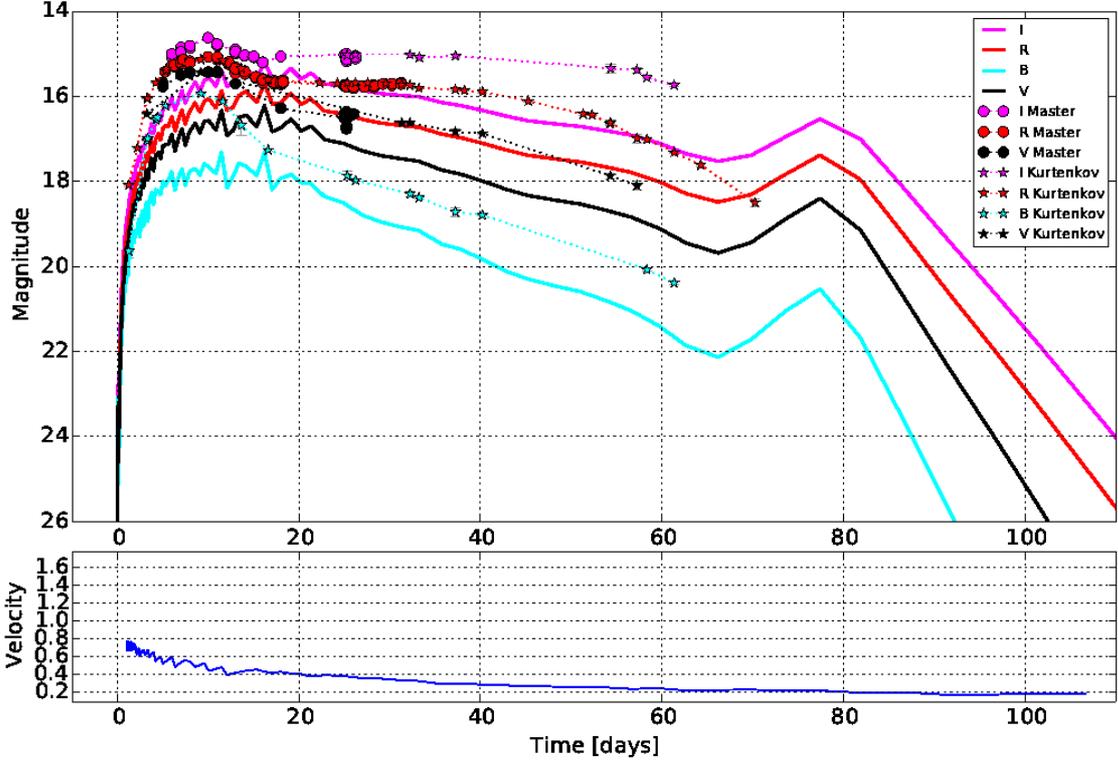}
  \caption{Light curves of the broadband photometry for the model run \texttt{R10\_M3Mht02t3\_E004} with fast heating in central layers.
  The secondary maximum occurs when the ejecta become transparent and the central
  hot remnant of the merged binary is visible.
  The lower panel shows the velocity on the level of photosphere for this model.}
\label{fig:ubvvel_rn90_R10_M3Mht02t3_E004wAq2e3}
\end{figure}

The nature of LRNs is different from that of collapsing supernovae in many aspects, that
is why we do not aim to reproduce all the details of observed light curves for the whole
period of observations using the \stella\ code.
This task would require developing of a new code and we leave this for future.
More modest problem is being solved in the current investigation: we try to elucidate
the physics of LRN emission on the plateau phase of the light curve, when it behaves
in a similar way to SN~IIP.
It is  the longest stage in the evolution of LRN with a characteristic behaviour of the
light curves, determined by the passage of cooling and recombination waves through the
expanding envelope.

The initial system we considered consists of two components: the inner core and and the outer shell.
Details of the inner core are not taken into account in our simulations, and the core
is treated as a hard sphere of a given mass.
The outer shell prior to the explosion is built as a model of a polytropic star, similar
to the one described in the article \citep{Baklanov2005}.
It is assumed that active dynamic processes in the binary and  the formation of the common envelope lead to a strong mixing of matter in the shell.
Therefore, the chemical composition in our simulations is uniform with solar
abundances at each point.

In this paper we are not going into the details of the mechanism.
We explored different ways of energy release:
from the fast release of all  energy through the explosion near the centre within the
mass of
$0.1-0.2 M_{\odot} $ to a long warm-up of the whole body of the star.

Version with the fast central heating with a time-scale of energy release
$t_{\rm heat} \sim 10^3$~s is shown in Fig.~\ref{fig:ubvvel_rn90_R10_M3Mht02t3_E004wAq2e3},
it demonstrates common features of such a heating.
The local release of energy in the central core of $M=0.2\, M_\odot$ produces a shock wave reaching the outer edge of envelope in $\sim 3^h$.
The shock wave propels all matter in the envelope with velocity higher than
the parabolic one.
This leads to the expansion of the envelope like in a supernova~IIP.
In comparison with our best-fit model (see below) a significant fraction of the energy goes into kinetic energy, which leads to a weaker
heating of matter and  a dim light on the plateau stage.
After  $ t \sim 60^d$ the envelope becomes transparent and  the heated core shines through it, giving the second maximum on the light curves.
The second maximum is not observed for M31LRN, but occurs in similar objects, so it is visible from LRN~V838~Mon \citep{Tylenda2005}.

It should be noted that the light curves are sensitive to changes in the initial masses and radii of envelopes \citep{Litvinova1985}.
See results for three models in Figs

\begin{figure}
  \centering
  \includegraphics[width=\linewidth]{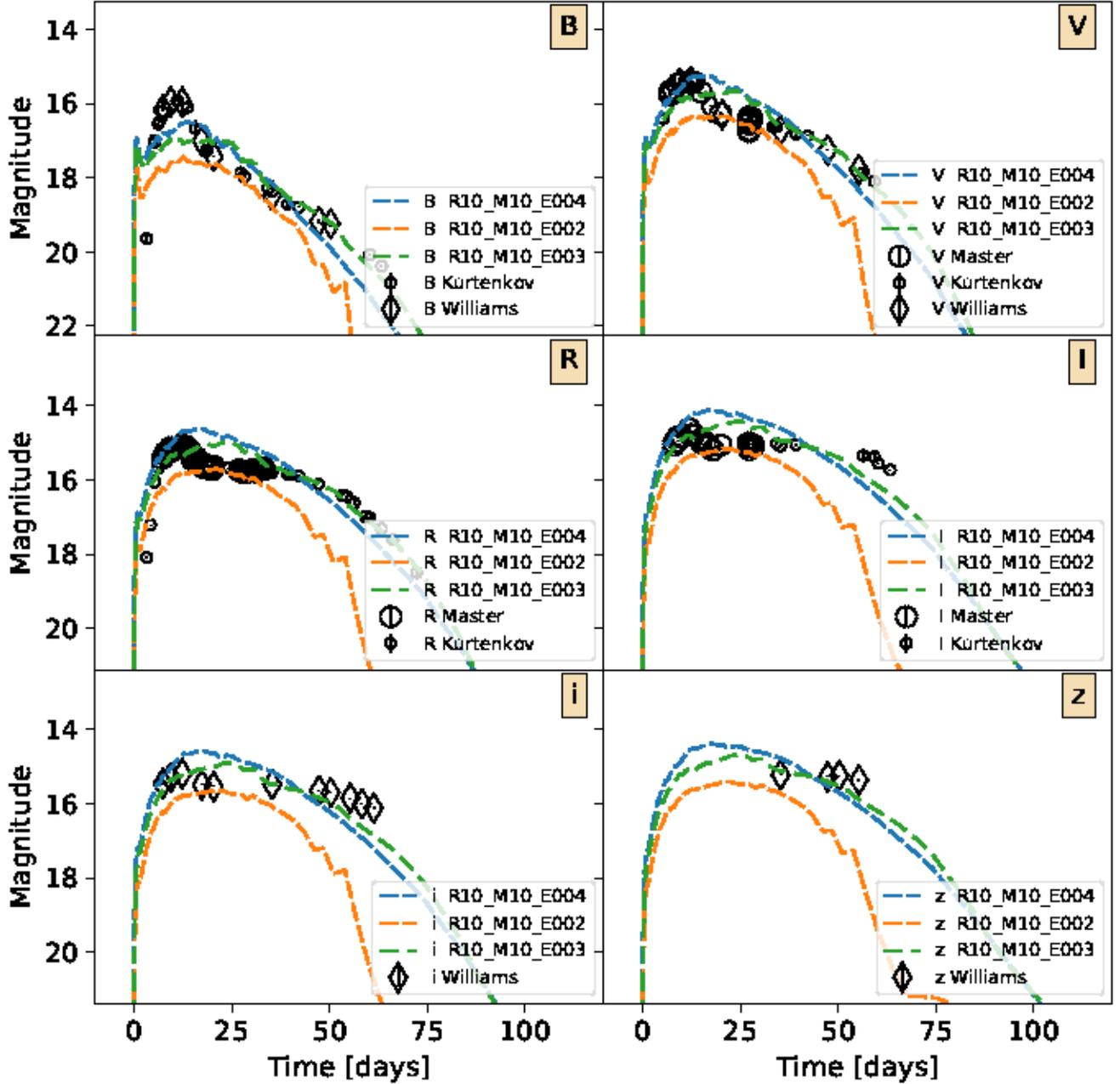}
  \caption{Light curves of the broadband photometry for the model runs  \texttt{lrnm31\_e00*} in various bands for variable energy.
  }
\label{fig:lrnm31_e002}
\end{figure}

\begin{figure}
  \centering
  \includegraphics[width=\linewidth]{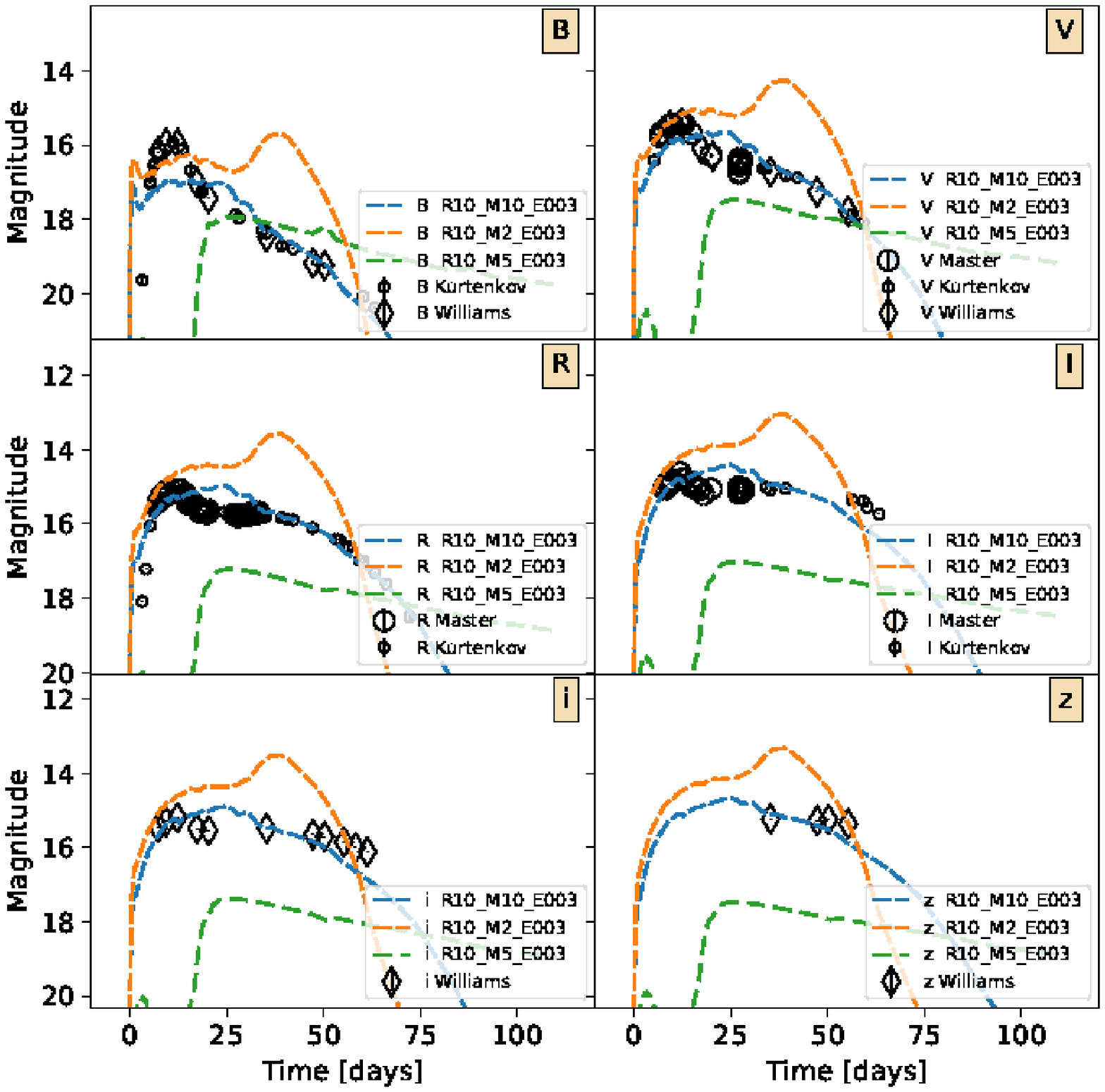}
  \caption{Light curves of the broadband photometry for the model run \texttt{lrnm31\_m*}
  in various bands for variable mass.
  }
\label{fig:lrnm31_m2}
\end{figure}

\begin{figure}
  \centering
  \includegraphics[width=\linewidth]{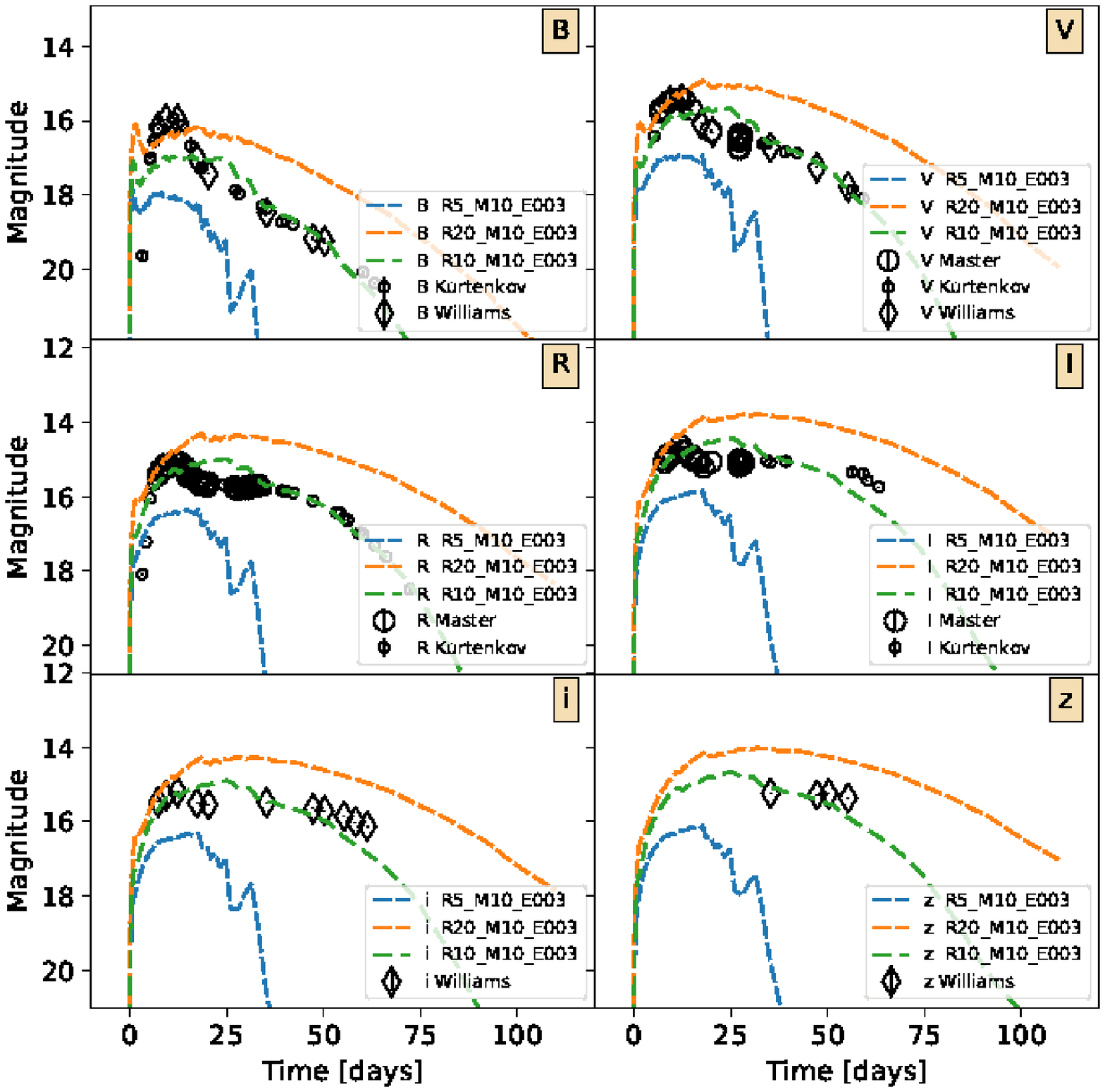}
  \caption{Light curves of the broadband photometry for the model run \texttt{lrnm31\_r*}
  in various bands for variable radius.
  }
\label{fig:lrnm31_r5}
\end{figure}

\begin{figure}
  \centering
  \includegraphics[width=\linewidth]{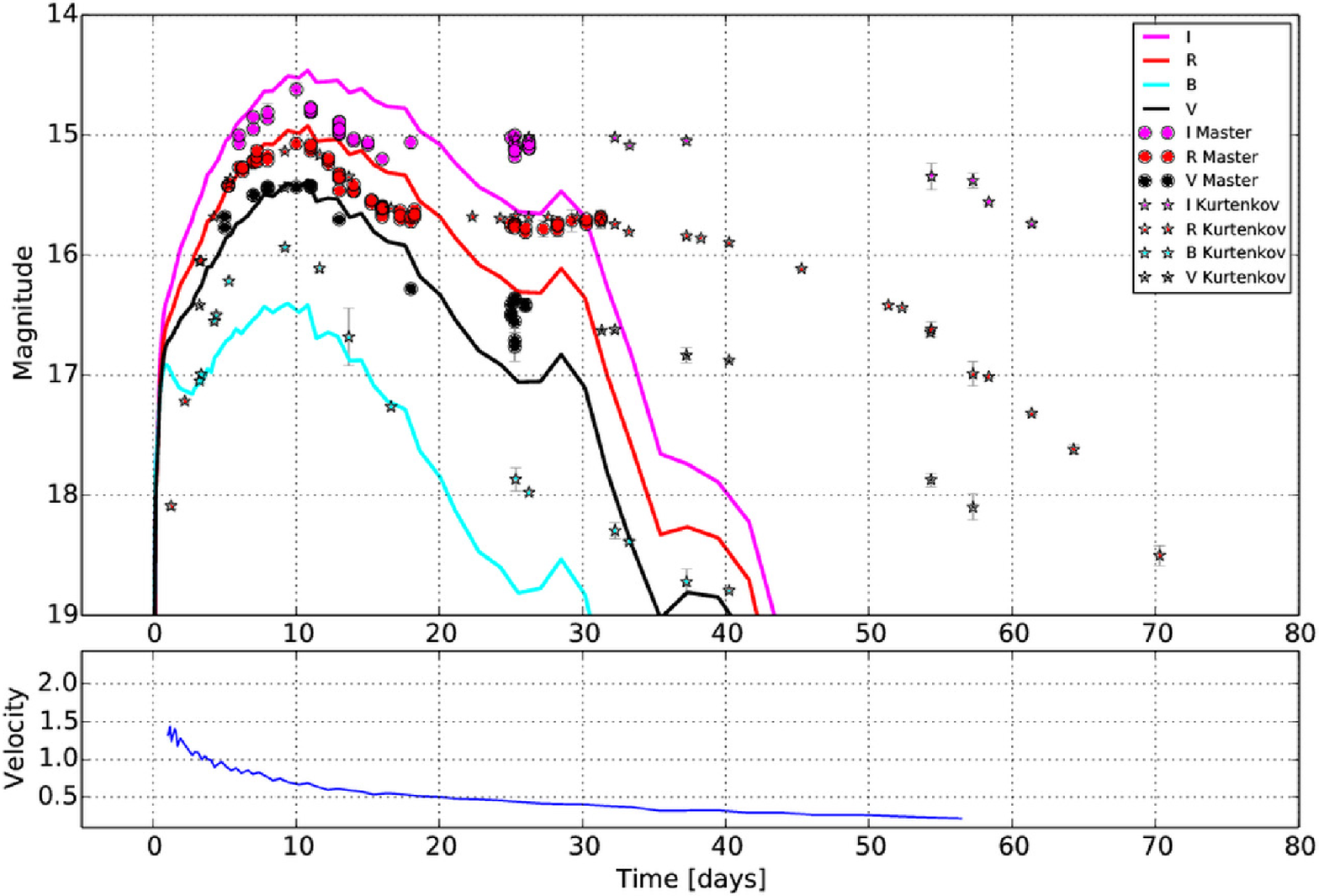}
  \caption{Light curves of broadband photometry for the model run demonstrating the absence of degeneracy in $R$ and $M$}
  \label{fig:ubvvel3}
\end{figure}

By varying  $R$, $M$ and $E$ one can achieve satisfactory matches of the model and
observed light curves at the maximum light, see  \ref{fig:ubvvel3}
For example, increasing the energy of the explosion to the $ E = 8 \times 10^{48} $~erg leads to a higher speed of the envelope expansion.
Light curves rise to the maximum faster, and the stage of CRW is shorter.
It can be seen that the model does not satisfactorily describe the observations since light curves do not match the observations after the maximum.
In addition, the physics of the light curve near the dome is not explained by a wave recombination \citep{MacLeod2016}.

\subsection*{Best-fit model}

The best fit to M31LRN observations is obtained with the model \mdlA, whose light curves are shown in Figure \ref{fig:ubvvel_rn300_R10_M3Mht3t30_Ni0_E003wAq2e3}.
The model has a total mass of $M_{\rm tot} = 3 M_{\odot}$ and the envelope radius
$ R = 10 R_{\odot} $.
The initial configuration was the same as the above-mentioned ``fast'' model.
The main difference between the two models is in the mode of energy release during
initiation  of LRN explosion.

Thermal energy  $E=3 \times 10^{48}$~erg is released throughout the whole mass
$M_{\rm tot}$ during a longer time $t_{\rm heat}\sim 10^4$~s.

It is evident that at the stage of the plateau, this model is in much better with observations.
The deviation in the band I, is apparently due to the insufficiently precise description of opacity in the infrared region.
When calculating the opacity in the lines we use a list of 150 thousand atomic transitions.
The main emphasis in the formation of lines in this list has been made on the ultraviolet and visible range of the spectrum (which is important for supernovae), while in the infrared region there is some shortage of lines {which is planned to fill out in future calculations}.

\subsection{The dynamics of expansion}

The dynamics of the expansion and the accompanying processes are
illustrated in a series of snapshots shown in Figure~\ref{fig:swd_rn300_R10_M3Mht3_Ni0_E003wAq2e3}.
 By the end of the first day of the shock has heated up the envelope and
accelerated the matter at the level of the photosphere to the velocity
$ v \sim 900 $~km/s.
Later, the inner layers lose their momentum pushing the overlying layers,
as well as due to the attraction of the underlying matter.
On day 10 the system has stabilized: about $ M \sim 2 \, M_\odot $ halts and stays in a
bound state, and about $ M_{\rm ej} \sim 1 \, M_\odot $ is ejected and
enteres the phase of free expansion.

The behavior of the light curves can be monitored on the Rosseland opacity curve (Tau)
in Figure~\ref{fig:swd_rn300_R10_M3Mht3_Ni0_E003wAq2e3}.
It is worth noting that Tau has been specially designed for those plots in order
to allow the qualitative monitoring the location of the photosphere.
Code \stella\ solves transport equations for the frequency grid from 1~\AA\ up to
$ 5\cdot 10^4$~\AA, without imposing any restrictions on the shape of the distribution  intensity and opacity in the cells of the frequency grid.

One can clearly see that the investigated plateau stage,
which is determined by the passage of a Cooling and Recombination Wave
\citep[CRW]{Grassberg1971}, begins already in the first days after the explosion.
In the right column of Figure~\ref{fig:swd_rn300_R10_M3Mht3_Ni0_E003wAq2e3} the X-axis
shows the Lagrangian coordinate $M(R)$.
It is evident that the CRW runs inside along the mass coordinate,
reaching the inner  stalled core at $ t \sim 90^d$.
According to the radial Eulerian coordinate (left column of
Figure~\ref{fig:swd_rn300_R10_M3Mht3_Ni0_E003wAq2e3}) the CRW is permanently
carried on by the expanding matter.
Therefore, the photospheric radius grows initially, providing a maximum of the light
curve at $\sim 25^d $, and then decreases slowly, reproducing the slow  decline in the
observed M31LRN bands.

\begin{figure}
  \centering
  \includegraphics[width=\linewidth]{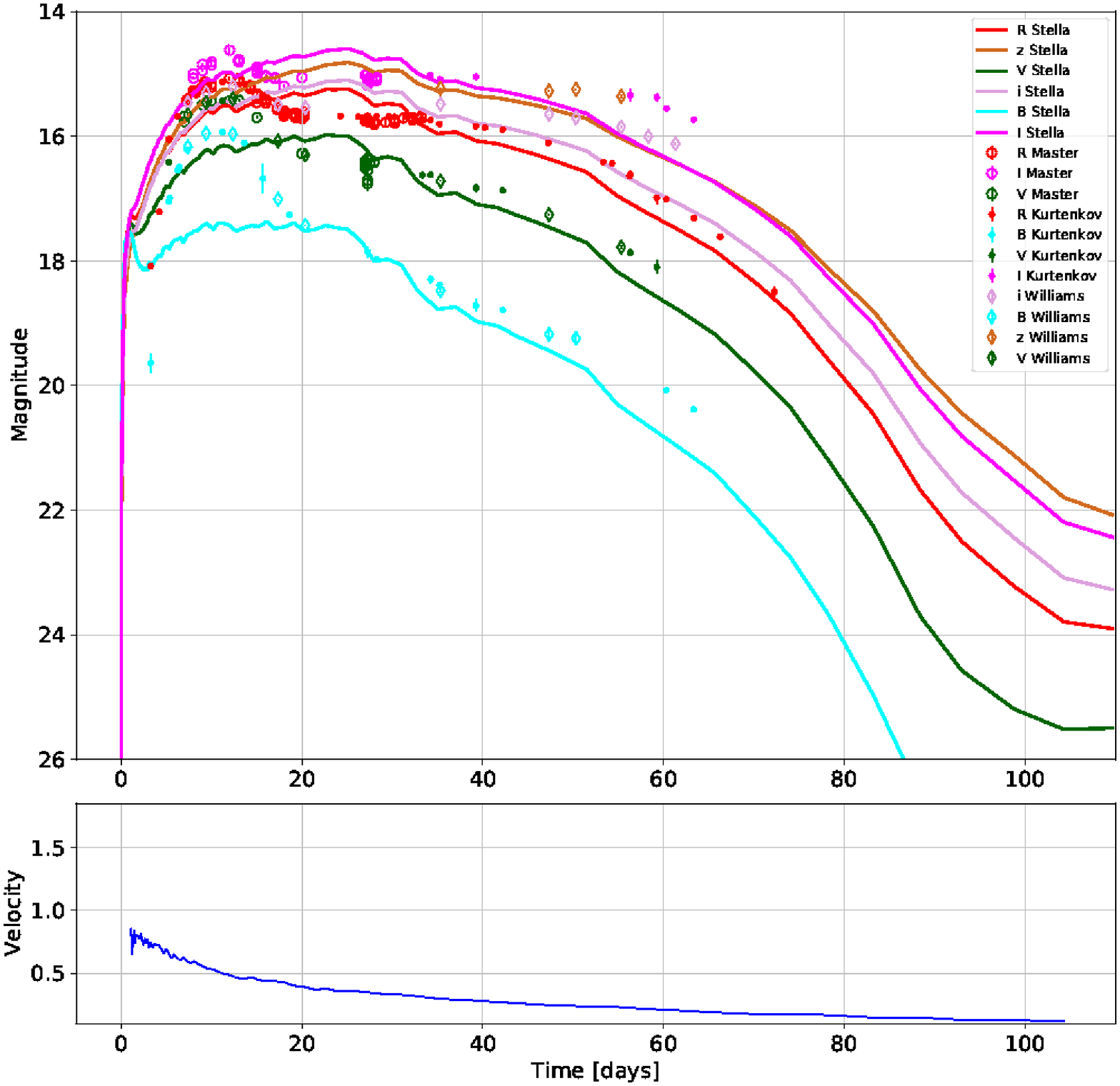}
  \caption{Light curves of broadband photometry for the model run \mdlA.
    The lower panel shows the speed at photosphere level in this model.
    This is the best model for the plateau stage.} \label{fig:ubvvel_rn300_R10_M3Mht3t30_Ni0_E003wAq2e3}
\end{figure}


\begin{figure}
  \centering
  \includegraphics[width=\linewidth]{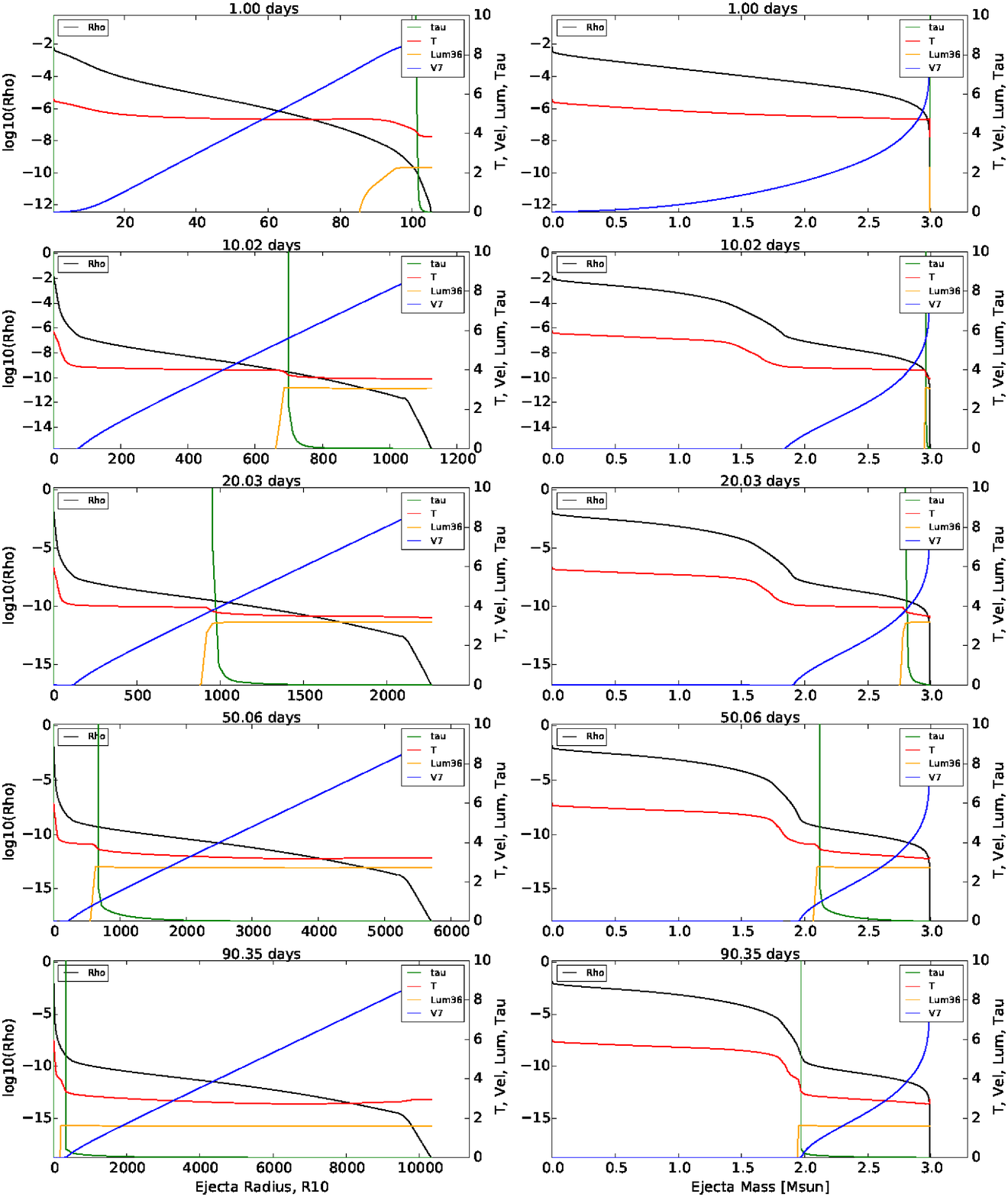}
  \caption{
LRN shell structure and its evolution as calculated in the model \mdlA at 1, 10, 20, 50, 90 days.
The color codes the graph of a physical quantity dependence on the radial variable:
density -- black [logarithmic scale, in units of g/cm$^3$]
bolometric luminosity -- orange [in units of $ 10^{36} $~erg/s],
temperature -- red [logarithmic scale, K],
velocity of matter -- blue [in units of $ 10^7 $~cm/s], Rosseland opacity -- green
[cm$^2$/g].
The difference between the left and right columns is the radial variable, plotted
along the axis X.
In the left column in the X-axis indicates a radius in units of $R_{sun}$,
and in the right hand column on the X axis using Lagrangean coordinate $M$ -- mass of the shell at a given radius.
}
 \label{fig:swd_rn300_R10_M3Mht3_Ni0_E003wAq2e3}
\end{figure}

It is important for the model not only to be consistent with photometric data but also to have hydrodynamic properties similar to those actually observed.
Figure~\ref{fig:ubvvel_rn300_R10_M3Mht3t30_Ni0_E003wAq2e3} shows the computed mass flow velocity at the photosphere level for the best-fit model \mdlA.
We have no detailed spectroscopic observations to superimpose onto our model computations.
 ~\citep{Williams2015} report $H\alpha$ data (FWHM = 900 $\pm$ 200 $km/s$), which provide an upper constraint for the photospheric velocity because the strong $H\alpha$ line forms above the photosphere, where the mass expansion velocity is higher than the photospheric velocity.
The photospheric velocity is better to estimate by weak lines.
In this case the value inferred from the NaI~D line (FWHM = 600 $\pm$ 200 $km/s$) agrees well with the model computations at the time of maximum light (fig.~\ref{fig:ubvvel_rn300_R10_M3Mht3t30_Ni0_E003wAq2e3}).

%


Our computations show that the overall behavior of broad-band light curves (Fig.~\ref{fig:ubvvel_rn300_R10_M3Mht3t30_Ni0_E003wAq2e3}) can be reproduced fairly well by codenamed model $\textbf{\mdlA}$.


We performed our computations for a spherically symmetric configuration consisting of an envelope and an inner core where the components of the system merge. In this formulation the component merger is treated as a source of thermal energy without taking into account the actual physics of the process
~\citep{Soker2003}.
Without performing a full multidimensional computation we cannot claim that our inferred total mass of\textbf{ $M_{\rm tot} =3 M_{\odot}$} characterizes the mass of the binary before the merger.
It is possible that in the case of a multidimensional computation the ejected mass and kinetic energy of the ejecta could be obtained with a less massive binary.

Williams et al. report the absolute magnitude of  $M_V=-1.5$ for the pre-supernova ~\citep{Williams2015}.
Main-sequence stars of such luminosity have the mass of $~6M_\odot$.
However, this estimate is not very reliable because it is unlikely that both merging objects are main-sequence stars. They are more probably red giants whose masses cannot be determined solely from their luminosity.
%

\subsection{Progenitor and evolution discussion}
One of the key evolutionary stages of binary stars is the specific state when the sizes of stars become comparable to that of the entire binary system ~\citep{Snezhko1968,Paczynski1966,vandenHeuvel1970,Svechnikov1974,Tutukov1981}

 It is a key stage because at that time violent mass exchange begins in the binary, resulting in dramatic and fast change of its parameters due to the exchange (and even loss) of the angular momentum and mass contained in the stars before the contact. The catastrophic change of binary parameters may result in complete merger of its component stars or in the abrupt decrease of the size of the binary semiaxis, and have a crucial effect on subsequent evolution stages that end with the formation of close binaries with degenerate components - neutron stars and white dwarfs. This is how white-dwarf close binaries form whose mergers may produce type Ia supernovas and, in the case of systems consisting of two neutron stars, short gamma-ray bursts. Hence mergers of two classical main-sequence stars or subgiants may contribute to the study and understanding of such important astrophysical phenomena as dark energy and gamma-ray bursts.

In the paradigm of synchronously rotating stars with a high mass ratio moving in circular
orbits the filling of the Roche lobe is followed by violent mass transfer by the primary
to the secondary and the formation of a common envelope. On the other hand, an equal-mass
binary may evolve into a fully contact system of two stars both of which simultaneously
fill their Roche lobes. In the limiting case violent mass transfer may occur on the
hydrodynamic time scale with the amount of released energy on the order of the thermal
energy of the component stars $\sim GM_{\rm tot}^2/ a \sim 8\cdot 10^{50} \mbox{erg/s} \cdot m^2 (a/R_\odot)$,
where $M_{tot}$
is the total mass of
colliding stars, and $a$ is the semimajor axis of the binary.

From this point of view it is very important to obtain observational evidence for such
processes in binary systems and to determine the corresponding physical properties - the
released energy, ejected mass, and angular momentum. Ultra-luminous red novae may be
observational manifestations of violent mass transfer in the binary \citep{Blagorodnova2017}.

We already pointed out that the observation of the progenitor of V1309~Sco in the form of a
contact binary with a period of 1.7 days \citep{Tylenda2011} provides a strong evidence for the
merger model of LRNe. This conclusion is also corroborated by the approximate
occurrence rate of such events -- one event in 20-40 years per galaxy like Milky Way or M31. To
show this, let us use the distribution of binary semimajor axes $a$ \citep{Tutukov1981}.
This distribution has the form
$dN  \sim 0.2 d \log (a/R_{\odot})$.
Systems with semi-axes  $6  \lesssim a/R_{\odot}  \lesssim 12$ merge during the Hubble time.
However, nova luminosities depend essentially on the mass of the ejected envelope, and the
occurrence rate of LRNe
may be several times lower, which is consistent
with observations of the M31 galaxy, where the last such nova was observed in 1989.

Or in another way, if all stars are close binaries then the maximum merger rate of systems with a total mass of 6 solar masses is equal to the birth rate of $3-6 M_\odot$ stars, which in the case of the Salpeter mass function is equal to ~ 1/10 year.
Given that close binaries make up for about 20-30 percent of all stars, we obtain an estimate of $1/30 - 1/50$ year. The last red nova in the Andromeda galaxy exploded 20 years ago, which is consistent with theoretical expectations

\section{Conclusions}
In this paper we described the technique of the discovery of the nova M31LRN and long-term observations of its light curve with MASTER network of robotic telescopes. It is important that the entire observational part of the study was performed on identical telescopes equipped with identical photometers. The resulting light curve agrees fairly well with the independent light curve published by ~\citep{Kurtenkov2015b}. However, our interpretation led us to infer a relatively higher total progenitor mass. The rather long plateau (~50 days) requires a higher merged stellar mass (~3 solar masses).
The corresponding explosion energy should be lower $2.5\cdot 10^{50}$~erg
,
whereas the total kinetic energy of the ejected envelope is lower by three orders of magnitude. The proposed interpretation of the explosion is consistent with the proposed evolutionary scenario where star merger is a natural stage of the evolution of close-mass stars and may serve as an extra channel for the formation of nova outbursts.


\section*{Acknowledgments}
MASTER Global Robotic Net is supported in part by the  Development Programm of Lomonosov Moscow State University.
This work was also supported in part by the RFBR grant 15-02-07875 (discovery and observations), by Russian Science Foundation grant 16-12-00085 (interpretation and data analysis) and grant 16-12-10519 (theoretical modelling of the Nova done by P.B). Grant no. IZ73Z0-152485 SCOPES Swiss National Science Foundation supports work of S.B.

\begin{table*}
 \centering
  \caption{ The R-band MASTER photometry of M31 2015 from MASTER-Kislovodsk and MASTER-Tunka observatories}
  \begin{tabular}{@{}lll@{}}
  \multicolumn{3}{|c|}{} \\
  \hline
The Data, MJD & R, mag & $R_{error}$ \\
 \hline
2457041.057  &  15.27  &  0.01\\
2457040.194  &  15.42  &  0.01\\
2457040.196  &  15.42  &  0.01\\
2457040.199  &  15.42  &  0.01\\
2457041.201  &  15.3  &  0.01\\
2457041.203  &  15.27  &  0.01\\
2457041.205  &  15.27  &  0.01\\
2457042.042  &  15.22  &  0.01\\
2457042.045  &  15.23  &  0.01\\
2457042.047  &  15.21  &  0.01\\
2457042.21  &  15.22  &  0.03\\
2457042.212  &  15.18  &  0.03\\
2457042.215  &  15.13  &  0.04\\
2457043.041  &  15.17  &  0.01\\
2457043.046  &  15.21  &  0.02\\
2457045.062  &  15.07  &  0.03\\
2457046.043  &  15.13  &  0.01\\
2457046.046  &  15.1  &  0.01\\
2457046.048  &  15.08  &  0.01\\
2457047.193  &  15.24  &  0.01\\
2457047.196  &  15.21  &  0.01\\
2457047.198  &  15.19  &  0.01\\
2457048.043  &  15.46  &  0.02\\
2457048.046  &  15.36  &  0.02\\
2457048.048  &  15.33  &  0.02\\
2457048.051  &  15.32  &  0.02\\
2457048.053  &  15.35  &  0.02\\
2457049.042  &  15.46  &  0.02\\
2457049.044  &  15.46  &  0.02\\
2457049.047  &  15.47  &  0.02\\
2457049.05  &  15.41  &  0.02\\
2457050.194  &  15.55  &  0.02\\
2457050.197  &  15.57  &  0.02\\
2457050.199  &  15.54  &  0.02\\
2457051.049  &  15.59  &  0.02\\
2457051.052  &  15.61  &  0.02\\
2457051.055  &  15.65  &  0.02\\
2457051.057  &  15.62  &  0.02\\
2457051.06  &  15.68  &  0.02\\
2457051.062  &  15.6  &  0.02\\
2457051.065  &  15.6  &  0.02\\
2457051.068  &  15.61  &  0.02\\
2457052.194  &  15.63  &  0.02\\
2457052.197  &  15.7  &  0.02\\
2457052.199  &  15.67  &  0.02\\
2457053.043  &  15.72  &  0.03\\
2457053.045  &  15.68  &  0.03\\
2457053.194  &  15.62  &  0.02\\
2457053.196  &  15.68  &  0.02\\
2457053.199  &  15.66  &  0.02\\
2457060.076  &  15.76  &  0.02\\
2457060.079  &  15.73  &  0.03\\
2457060.126  &  15.75  &  0.03\\
2457060.128  &  15.74  &  0.03\\
2457060.151  &  15.75  &  0.04\\
2457060.154  &  15.74  &  0.04\\
2457060.26  &  15.77  &  0.03\\
2457061.118  &  15.81  &  0.03\\
2457061.121  &  15.77  &  0.03\\
2457062.198  &  15.78  &  0.07\\
2457063.214  &  15.79  &  0.06\\
2457063.217  &  15.75  &  0.04\\
2457063.219  &  15.73  &  0.03\\
2457063.229  &  15.75  &  0.02\\
2457064.272  &  15.71  &  0.09\\
2457065.235  &  15.72  &  0.02\\
2457065.238  &  15.74  &  0.02\\
2457065.243  &  15.7  &  0.03\\
2457066.253  &  15.68  &  0.06\\
2457066.248  &  15.69  &  0.06 \\
2457066.251  &  15.71  &  0.07 \\
\hline
\end{tabular}
\label{TabGTCR}\\
\end{table*}

\begin{table*}
 \centering
  \caption{ The V-band MASTER photometry of M31 2015 from MASTER-Kislovodsk and MASTER-Tunka observatories}
  \begin{tabular}{@{}lll@{}}
    \multicolumn{3}{|c|}{} \\
  \hline
The Data, MJD & R, mag & $R_{error}$ \\
 \hline
2457040.117  &  15.68  &  0.03\\
2457040.122  &  15.77  &  0.03\\
2457042.045  &  15.5  &  0.01\\
2457042.047  &  15.5  &  0.01\\
2457043.041  &  15.44  &  0.01\\
2457043.043  &  15.45  &  0.02\\
2457043.046  &  15.43  &  0.02\\
2457045.062  &  15.43  &  0.03\\
2457046.043  &  15.43  &  0.01\\
2457046.046  &  15.43  &  0.01\\
2457046.048  &  15.42  &  0.02\\
2457048.053  &  15.7  &  0.02\\
2457053.045  &  16.28  &  0.04\\
2457060.076  &  16.5  &  0.04\\
2457060.077  &  16.49  &  0.06\\
2457060.079  &  16.41  &  0.06\\
2457060.14  &  16.37  &  0.07\\
2457060.143  &  16.71  &  0.1\\
2457060.148  &  16.55  &  0.09\\
2457060.151  &  16.36  &  0.08\\
2457060.154  &  16.76  &  0.12\\
2457061.118  &  16.42  &  0.05\\
2457061.121  &  16.41  &  0.05\\
\hline
\end{tabular}
\label{TabGTCV}\\
\end{table*}

\begin{table*}
 \centering
  \caption{ The I-band MASTER photometry of M31 2015 from MASTER-Kislovodsk and MASTER-Tunka observatories}
  \begin{tabular}{@{}lll@{}}
    \multicolumn{3}{|c|}{} \\
  \hline
The Data, MJD & R, mag & $R_{error}$ \\
 \hline
2457041.009  &  15.07  &  0.02\\
2457041.035  &  15  &  0.02\\
2457042.017  &  14.85  &  0.02\\
2457042.019  &  14.95  &  0.02\\
2457043.01  &  14.86  &  0.03\\
2457043.012  &  14.81  &  0.07\\
2457045.058  &  14.62  &  0.05\\
2457046.009  &  14.81  &  0.02\\
2457046.011  &  14.78  &  0.02\\
2457046.014  &  14.8  &  0.02\\
2457046.016  &  14.8  &  0.02\\
2457046.019  &  14.77  &  0.01\\
2457048.011  &  14.94  &  0.02\\
2457048.014  &  14.99  &  0.02\\
2457048.016  &  14.89  &  0.02\\
2457048.019  &  14.94  &  0.02\\
2457048.021  &  14.89  &  0.02\\
2457048.024  &  14.98  &  0.02\\
2457048.027  &  14.95  &  0.02\\
2457049.015  &  15.05  &  0.02\\
2457049.017  &  15.03  &  0.02\\
2457050.014  &  15.08  &  0.02\\
2457050.017  &  15.06  &  0.02\\
2457051.012  &  15.2  &  0.03\\
2457053.014  &  15.06  &  0.02\\
2457060.085  &  15.02  &  0.02\\
2457060.158  &  15.11  &  0.04\\
2457060.16  &  15.08  &  0.03\\
2457060.163  &  15  &  0.03\\
2457060.166  &  15.18  &  0.04\\
2457060.176  &  15.13  &  0.04\\
2457061.126  &  15.09  &  0.03\\
2457061.129  &  15.11  &  0.03\\
2457061.131  &  15.05  &  0.03\\
2457061.136  &  15.11  &  0.03\\
2457061.142  &  15.07  &  0.03\\
\hline
\end{tabular}
\label{TabGTCI}\\
\end{table*}

\begin{table*}
 \centering
  \caption{ MASTER unfiltered (W) photometry of M31 2015 from MASTER-Kislovodsk and MASTER-Tunka observatories}
  \begin{tabular}{@{}lll@{}}
    \multicolumn{3}{|c|}{} \\
  \hline
The Data, MJD & W, mag & $W_{error}$ \\
 \hline
2457036.134  &  18.27  &  0.05\\
2457036.137  &  18.43  &  0.06\\
2457036.139  &  18.3  &  0.05\\
2457037.124  &  17.4  &  0.06\\
2457037.127  &  17.46  &  0.04\\
2457037.129  &  17.66  &  0.04\\
2457038.127  &  16.44  &  0.02\\
2457038.129  &  16.43  &  0.02\\
2457039.137  &  15.88  &  0.01\\
2457039.139  &  15.91  &  0.01\\
2457039.141  &  15.89  &  0.01\\
2457040.129  &  15.6  &  0.01\\
2457040.131  &  15.57  &  0.01\\
2457040.179  &  15.61  &  0.01\\
2457040.181  &  15.65  &  0.01\\
2457040.183  &  15.6  &  0.01\\
2457041.127  &  15.43  &  0.01\\
2457041.129  &  15.43  &  0.01\\
2457041.132  &  15.43  &  0.01\\
2457041.196  &  15.46  &  0.01\\
2457041.198  &  15.45  &  0.01\\
2457042.13  &  15.39  &  0.02\\
2457042.133  &  15.39  &  0.02\\
2457042.207  &  15.38  &  0.02\\
2457043.236  &  15.35  &  0.01\\
2457043.239  &  15.35  &  0.01\\
2457043.241  &  15.35  &  0.01\\
2457044.194  &  15.34  &  0.02\\
2457044.196  &  15.29  &  0.02\\
2457045.246  &  15.28  &  0.02\\
2457046.148  &  15.33  &  0.02\\
2457047.132  &  15.4  &  0.01\\
2457047.135  &  15.39  &  0.01\\
2457047.137  &  15.41  &  0.01\\
2457047.182  &  15.43  &  0.01\\
2457047.184  &  15.42  &  0.01\\
\hline
\end{tabular}
\label{TabGTCW}\\
\end{table*}

\bibliographystyle{mnras}
\bibliography{redNova}

\bsp	
\label{lastpage}

\end{document}